\documentclass[transmag]{IEEEtran}
\usepackage[utf8]{inputenc}
\usepackage[square, numbers, sort&compress]{natbib}
\usepackage{graphicx}
\usepackage{xparse,nameref}
\usepackage[colorlinks=true]{hyperref}
\usepackage{caption} 

\usepackage{booktabs} 
\usepackage{xcolor}
\usepackage{amsmath, amssymb}
\usepackage{cleveref} 

\NewDocumentCommand{\citeref}{s m}{Ref.~\cite{#2}\IfBooleanF{#1}{ \nameref{#2}}} 
\NewDocumentCommand{\sectref}{s m}{Section~\ref{#2}\IfBooleanF{#1}{ \nameref{#2}}} 
\NewDocumentCommand{\figref}{s m}{Figure~\ref{#2}\IfBooleanF{#1}{ \nameref{#2}}} 
\NewDocumentCommand{\eqnref}{s m}{Eqn~\ref{#2}\IfBooleanF{#1}{ \nameref{#2}}} 
\NewDocumentCommand{\tblref}{s m}{Table~\ref{#2}\IfBooleanF{#1}{ \nameref{#2}}} 

\begin{document}

\title{Persistent Weak Interferer Detection for WiFi Networks: A Deep Learning Based Approach}

\IEEEtitleabstractindextext{%
    \begin{abstract}
\emph{Abstract}---In this paper, we explore the use of multiple deep learning techniques to detect weak interference in WiFi networks. Given the low interference signal levels involved, this scenario tends to be difficult to detect. However, even signal-to-interference ratios exceeding 20~dB can cause significant throughput degradation and latency. Furthermore, the resultant packet error rate may not be enough to force the WiFi network to fallback to a more robust physical layer configuration. Deep learning applied directly to sampled radio frequency data has the potential to perform detection much cheaper than successive interference cancellation, which is important for real-time persistent network monitoring. The techniques explored in this work include maximum softmax probability, distance metric learning, variational autoencoder, and autoreggressive log-likelihood. We also introduce the notion of generalized outlier exposure for these techniques, and show its importance in detecting weak interference. Our results indicate that with outlier exposure, maximum softmax probability, distance metric learning, and autoreggresive log-likelihood are capable of reliably detecting interference more than 20~dB below the 802.11 specified minimum sensitivity levels. We believe this presents a unique software solution to real-time, persistent network monitoring.   
\end{abstract}
}

\author{%
  \IEEEauthorblockN{%
  Andrew Adams, 
  Richard F. Obrecht, 
  Miller Wilt,\\ 
  Daniel Barcklow, 
  Bennett Blitz,
  Daniel Chew
  }%
  \\
  \IEEEauthorblockA{The Johns Hopkins University Applied Physics Laboratory}
}


\IEEEpubid{0000--0000/00\$00.00~\copyright~2022 IEEE}

\maketitle

\section{Introduction \label{sect:Intro}}
\IEEEPARstart{W}{ireless} connectivity has become ubiquitous in modern society.
This is illustrated by the vast amounts of internet traffic we produce. In 2020, global internet traffic reached an average of 175 exabytes per month, and is expected to reach 300 exabytes per month by 2022~\cite{internetstats2019}. This is an enormous amount of data being trafficked, which indicates the level of integration internet access has in our daily lives. Although our consumption of wireless services continues to increase, spectrum regulations have lagged in supporting the increased traffic rates~\cite{Chew-et-al-2021}. Nowhere is this more evident than the unlicensed Industrial, Scientific, and Medical (ISM) frequency bands, where WiFi is forced to operate. Here WiFi competes with itself (co-located networks) and other popular wireless standards for spectrum access. These include Long Term Evolution License Assisted Access (LTE-LAA),  Bluetooth, ZigBee, cordless phones, radio frequency identification (RFID) tags, and many proprietary wireless devices and protocols. As such, the ability to detect and mitigate sources of interference is central to any successful WiFi network deployment.  

\textit{Interference} occurs when WiFi transmissions collide with other traffic attempting to use the same or adjacent overlapping channels. This induces various signal processing errors within the intended receiver resulting in reduced network throughput and increased network latency. As discussed in \citeref*{Gummadi-et-al-2007}, even relatively weak levels of interference can severely degrade WiFi network performance. The authors demonstrate a significant reduction in throughput and an increase in latency with signal-to-interferer (SIR) ratios as large as 20~dB. However, these can be the most difficult to detect from a network monitoring perspective, whereby the necessary mitigation strategy is never initiated.

In this paper, we explore multiple deep learning (DL) methods to enable persistent detection of weak interferers. Our motivation to explore DL for this application is twofold. First, detection of weak interferers enables network administrators to execute appropriate mitigation strategies, either manually or autonomously, with the most effective being frequency agility~\cite{Gummadi-et-al-2007}. Furthermore, given the consistent use of WiFi to facilitate internet traffic, a persistent monitoring capability can run 24/7, and alert network administrators only when mitigation is required. 

Second, DL has the potential to provide a persistent monitoring capability at a fraction of the computational cost of existing techniques, including successive interference cancellation (SIC)~\cite{Souvik2010}. When employing SIC, any unwanted traffic is first demodulated to determine content, re-modulated, and then subtracted from the input sample stream. The resulting residue is then analyzed for indications of desired signal traffic or potential signals of interest (SOI). This is a highly complex process, where the performance depends heavily upon accurate estimates of both hardware and channel impairments by the intended receiver. This process is discussed at length in \citeref*{Selim2019}, where in-phase/quadrature (IQ) imbalance, direct current (DC) offset, and power amplifier (PA) non-linearities are shown to reduce the performance of non-orthogonal multiple-access (NOMA) communications systems. Another example is the bit error rate (BER) floor induced by uncompensated transmitter windowing in orthogonal frequency division multiplexing (OFDM) communications systems~\cite{Huang2001}. 

Given these motivating factors, we demonstrate the use of select DL models to detect weak interferers directly 
from the received sample stream without the need for SIC or any other unwanted signal removal or impairment compensation.  
\IEEEpubidadjcol

\section{Experiment Setup \label{sect:ExpSetup}}
\begin{figure}[t!]
\begin{center}
\includegraphics[scale=0.485]{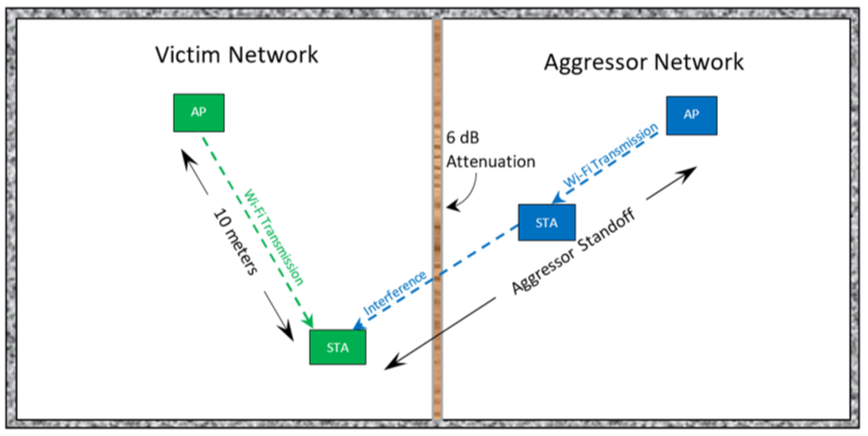}
 \caption[]{An example of a co-channel interference scenario.}
 \label{fig:Adj_Nwk_CCI}
\end{center}
\end{figure}

For experimentation purposes, we chose to replicate the common occurrence of co-channel interference caused by frequency 
reuse of wireless local area network (WLAN) channels between co-located WiFi networks. This scenario is illustrated 
in \figref*{fig:Adj_Nwk_CCI}, where two simple networks, each consisting of a single access point (AP) and wireless 
station (STA), are separated geographically but attempt traffic exchange on the same WLAN channel. The \textit{victim} 
network is so named because in this scenario it began traffic exchange on a particular WLAN channel first; the 
\textit{aggressor} network is so named because in this scenario it began traffic exchange on the same WLAN channel 
at some point thereafter. Victim and aggressor roles hold relative meaning only, and could be reversed at any time.

\begin{figure}[t!]
\begin{center}
\includegraphics[scale=0.5]{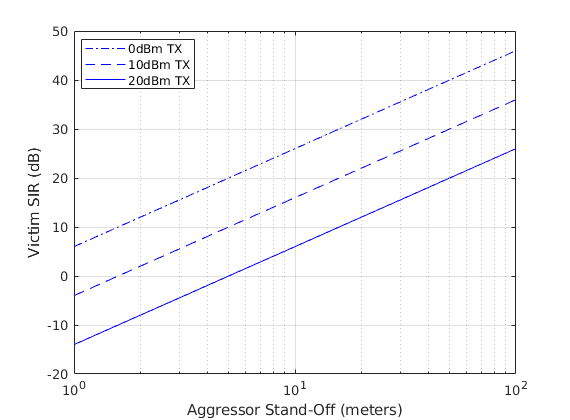}
 \caption[]{Victim STA interference levels as a function of aggressor distance.}
 \label{fig:Victim_Int_Lvls}
\end{center}
\end{figure}

In this scenario, the aggressor network operates in close proximity, such as an adjacent office space or outdoor
block. As the aggressor network begins traffic exchange, the victim STA experiences co-channel interference; the
magnitude of  which depends upon the aggressor STA transmit power, standoff, and signal attenuation due to
intervening walls or  partitions. Assuming a victim AP power level of +20~dBm, aggressor STA transmit power levels
ranging from 0~dBm to +20~dBm, aggressor STA standoffs ranging from 1~m to 100~m, and 6~dB partition loss,
\figref*{fig:Victim_Int_Lvls} shows the  resultant signal-to-interference ratios (SIR) at the victim STA. These
levels suggest the victim STA packet error rates (PER) may not exceed 10$\%$, which means the co-channel 
interference may not be enough to force it to a lower modulation coding scheme (MCS)~\cite{80211-2020}. 
However, as demonstrated in \citeref*{Gummadi-et-al-2007}, it is certainly enough to reduce victim network
throughput and quality of service (QoS). Hence our desire to pursue  detection methods relevant to \textit{weak}
interferers.

We therefore explore the ability of select DL techniques to detect weak interferers for various static MCS 
values. Assuming a victim STA OFDM physical layer (PHY) configuration, \citeref*{80211-2020} specifies the 
minimum required receiver sensitivity for each MCS value. Assuming the noise figure and implementation loss 
specified there, room temperature, and a noise bandwidth of 20~MHz, we can derive the minimum specified operating
signal-to-noise ratio, SNR$_{\textrm{min}}$(dB), for each subcarrier modulation type specified. These include binary
phase shift keying (BPSK), quadrature phase shift keying (QPSK), and quadrature amplitude modulation (QAM) as 
listed in \tblref*{tab:Mod_SINR}. For our experiments, SNR and SIR are combined into a metric termed
\textbf{signal-to-interference-and-noise ratio} (SINR). Since the interference level is not enough to force the
victim STA to a lower MCS value, the SINR is assumed to include an SIR range between  SNR$_{\textrm{min}}$(dB)
and SNR$_{\textrm{min}}-20$~dB when present. This is expressed mathematically as 
\begin{align}
  \textrm{SINR} &= 10^{-\textrm{SNR}_{\textrm{min}} / 10} \nonumber \\
                &= 10^{-\textrm{SIR} / 10} + 10^{-\textrm{SNR} / 10},
\end{align}
where SNR$_{\textrm{min}}$ is defined in \tblref*{tab:Mod_SINR}. SIR is drawn from a uniform distribution 
$\in [10^\frac{-\textrm{SNR}_{\textrm{min}}-20}{10}, 10^\frac{-\textrm{SNR}_{\textrm{min}}}{10}]$, and SNR 
consumes the difference. The result is a range of continuous SINR values dominated by SIR or SNR on either 
extreme. The outcome of each experiment is thus the quantitative measure of our ability to use DL to accurately 
detect co-channel intereference at a given SIR when present.

\begin{table}[b!]
\caption{Minimum SINR per subcarrier modulation type.}
\centering
    \begin{tabular}{l r} \toprule
        Modulation &  SNR$_{\textrm{min}}$~(dB) \\ \midrule
        BPSK   & 3.92   \\
        QPSK   & 6.92   \\
        16-QAM & 11.92  \\
        64-QAM & 19.92  \\ \bottomrule
        \hline
    \end{tabular}
\label{tab:Mod_SINR}
\end{table}

\section{Relevant Works \label{sect:Relevant_Works}}
To our knowledge, our DL based approach to weak interferer detection is novel. However, relevant performance and complexity comparisons can be made to multiple access schemes employing SIC. Briefly, SIC allows multiple transmitters to occupy a common channel simultaneously. At the receiver, any unwanted traffic is first demodulated, re-modulated, and then subtracted from the input sample stream. To do so, the desired traffic is treated as co-channel interference while demodulating the undesired traffic; hence our basis for comparison. Finally, the desired traffic is then recovered from the resulting residue when present. For packetized transmissions, this is done in \textit{succession} to recover the entire message, hence the term \textit{successive interference cancellation}. We do not present a quantitative analysis of relative complexity here, as that entails algorithmic implementation details for either approach. We do however highlight relevant works discussing current perceived limitations of SIC as motivation to explore other approaches. 

\citeref*{Souvik2010} discusses the medium access control (MAC) layer throughput gains provided by SIC as compared to a time sharing approach such as time division multiple access (TDMA). Their conclusion is that throughput gains are marginal for distinct transmitter-receiver pairs. That is, as a means to combat co-channel interference, SIC provides little to no benefit. The reason being that for SIC to be feasible, certain requirements regarding transmitter bit rates and signal strength must be met. For multiple transmitters and a common receiver, SIC then provides modest gains as the transmitters can coordinate to meet said feasibility requirements. 

\citeref*{Souvik2010} assumes perfect cancellation, which is of course unobtainable. \citeref*{Selim2019} illuminates this challenge while discussing performance degradation within NOMA schemes due to RF front-end impairments; NOMA relies heavily upon SIC to eliminate multi-user interference. To minimize receiver complexity, impairments such as IQ imbalance, DC offset, and PA non-linearities are often ignored. While these often have little effect on TDMA schemes, these present significant performance degradation to NOMA schemes in the form of imperfect cancellation. \citeref*{Huang2001} discusses a similar concern with OFDM windowing techniques, which are often not standardized or known a priori by the intended receiver. 

\section{Deep Learning Based Approaches \label{sect:DL_Approaches}}
In this work, we demonstrate detection of weak interferers using various deep learning approaches. Deep learning is a
form of representation learning, where complex concepts are built from combinations of simpler 
ones~\cite{Goodfellow-et-al-2016}. These are commonly implemented using artificial neural networks with numerous 
hidden layers, \emph{i.e.} deep, where each layer tends to learn different features from the preceding layers. This results
in a specific concept being learned by iteratively traversing the network, depth-wise, as part of an optimization 
algorithm, such as stochastic gradient descent. Once trained, the inferred network approximates a function
that produces the desired output from a given input, and can be expressed mathematically as $y = f(x)$,
where $f(x)$ is not limited to linear functions. Deep learning has been shown adept in performing various learning 
tasks across many disciplines and applications. Examples related to wireless communications include modulation
recognition directly from baseband samples~\cite{oshea2016convolutional,Karra2017ModulationRU,Rajendran2018DeepLM} 
and reinforcement learning Q-function
approximation for network resource scheduling~\cite{Chang2019, Tathe2018}. We experiment with deep anomaly 
detection as a means to detect weak co-channel interference within WiFi networks. As such, we train neural networks to 
produce multiple feature representations and anomaly scores indicative of situations where weak interferers may be reducing
network throughput or inducing additional traffic latency. 

For the interested reader, \citeref*{Pang_2021} provides a comprehensive review of deep anomaly detection methods. We 
instead focus on a closely related form of deep anomaly detection called \textbf{out-of-distribution} (OOD) detection,
which is attractive for this application because it enables anomaly detection in an open-world scenario; that 
is, we assume we have sufficient labeled examples of what normal looks like, but make no assumptions with respect to 
the types of anomalies encountered once deployed. As such, \textbf{in-distribution} (ID) refers to the traffic exchanged
between nodes within our own network without any weak interferers present; OOD then refers to similar traffic exchange,
but in the presence of weak co-channel interference. 

\begin{figure}[t!]
\begin{center}
\includegraphics[scale=0.18]{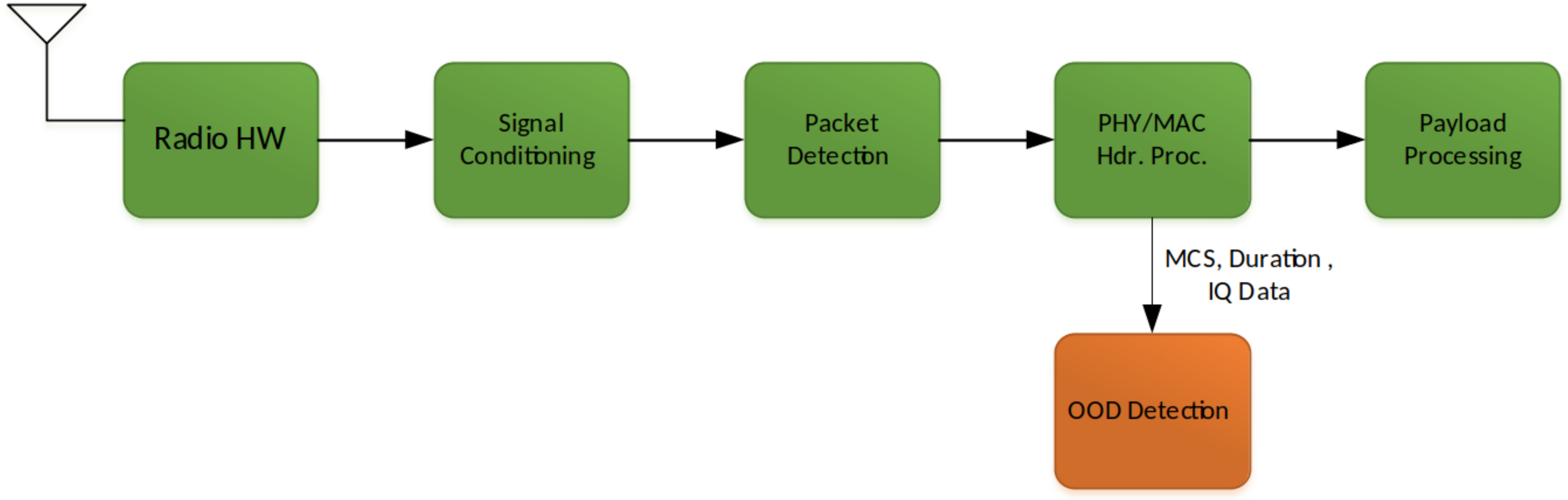}
 \caption[]{An OOD detection application.}
 \label{fig:OOD}
\end{center}
\end{figure}

Since we are processing baseband (IQ) samples directly, normal necessarily depends upon the various network physical 
layer configurations. Since this is an in-network monitoring application, we assume we can detect each packet
and recover the PHY and MAC headers such that the MCS and packet duration are known prior to network inference. This 
reduces the scope of our experiments to OOD detection, and prevents the detection algorithms from processing noise only.
\figref*{fig:OOD} illustrates how this application may be deployed alongside regular traffic processing on a given 
observation node or AP.  

Note that other published works, including Refs.~\cite{Pang_2021, hendrycks2018baseline, hendrycks2019deep},
distinguish OOD detection from other anomaly detection methods by the capability to do so alongside primary learning 
tasks such as classification. While we believe this to be beneficial, it is not the focus of our experimentation. 
However, future work may include this capability to detect weak interferers with little to no header
information. This would certainly be attractive when attempting to monitor WiFi networks without first joining the network
being monitored.

\subsection{Out of Distribution Detection}
OOD refers to examples drawn from a different distribution than that which are considered to be ID or normal~\cite{Pang_2021,
hendrycks2018baseline, hendrycks2019deep, techapanurak2021practical}. Detecting OOD examples can be done by training deep neural networks to
generate various representations or metrics of the input data, which adequately differ between OOD and ID examples. However,
for OOD examples to be detected within an open-world deployment scenario, we can make no assumptions regarding the types of 
anomalies during network inference. Since that is our goal here, we focus on (4) specific methods of OOD, assuming only the
availability of labeled ID training data: \textbf{maximum softmax probabilities~(MSP)}, 
\textbf{distance metric learning~(DML)} similarity scores, 
\textbf{variational autoencoder~(VAE)} reconstruction loss, and \textbf{autoregressive~(AR)} model likelihood.
We elaborate on each method following an introduction to outlier exposure. 

\subsection{Outlier Exposure} 
\subsubsection{Motivation} \label{subsect:OE}
While many OOD detection models detect anomalies using learned representations of ID examples only, 
an auxiliary OOD dataset may be used to learn heuristics of OOD examples, a technique known as outlier exposure
(OE)~\cite{hendrycks2019deep}. This is motivated by the use of complex datasets, where anomalies are
difficult to detect, which is particularly of concern here given the diversity of WiFi signal formats.
Hendrycks \emph{et al.} represent outliers using disjoint datasets $\mathcal{D}_{\textrm{in}}$ and
$\mathcal{D}_{\textrm{out}}^{\textrm{OE}}$; that method is not applicable here as each anomalous example
still represents a valid class. In fact, one may find parallels between this application and the notion 
of domain shift as discussed in \citeref*{techapanurak2021practical}. Unfortunately, we do not have the 
same benefit of using multiple examples to detect weak interferers as that has the potential to introduce
significant latency into our monitoring system. We still utilize OE to train our networks to be sensitive to the 
presence of weak interferers; however, our  outliers must be representative of relatively weak co-channel
interference.

Given $x$ drawn from an ID dataset $\mathcal{D}_{\textrm{in}}$, a class label $y$, $x'$ drawn from an OOD dataset 
$\mathcal{D}_{\textrm{out}}^{\textrm{OE}}$, a model $f$, and the original learning objective $\mathcal{L}$, outlier
exposure is defined by minimizing the following objective:
\begin{align}
    \mathbb{E}_{(x,y) \sim \mathcal{D}_{\textrm{in}}}
        \left[ 
             \mathcal{L} \left(f(x), y\right) + 
             \lambda \mathbb{E}_{x' \sim \mathcal{D}_{\textrm{out}}^{\textrm{OE}}} 
                    \left[ \mathcal{L}_{\textrm{OE}} \left(f(x'), f(x), y \right) \right]
        \right],
    \label{eqn:OE_loss}
\end{align}
where $\lambda$ is a scaling factor and $\mathcal{L}_{\textrm{OE}}$ is a function of $f(x')$, $f(x)$, and $y$. 
In other words, the process is to first maximize the classification accuracy using data drawn from $\mathcal{D}_{\textrm{in}}$,
and then to minimize confidence using data drawn from $\mathcal{D}_{\textrm{out}}^{\textrm{OE}}$. 

\begin{figure}[t!]
    \begin{center}
        \includegraphics[scale=0.3]{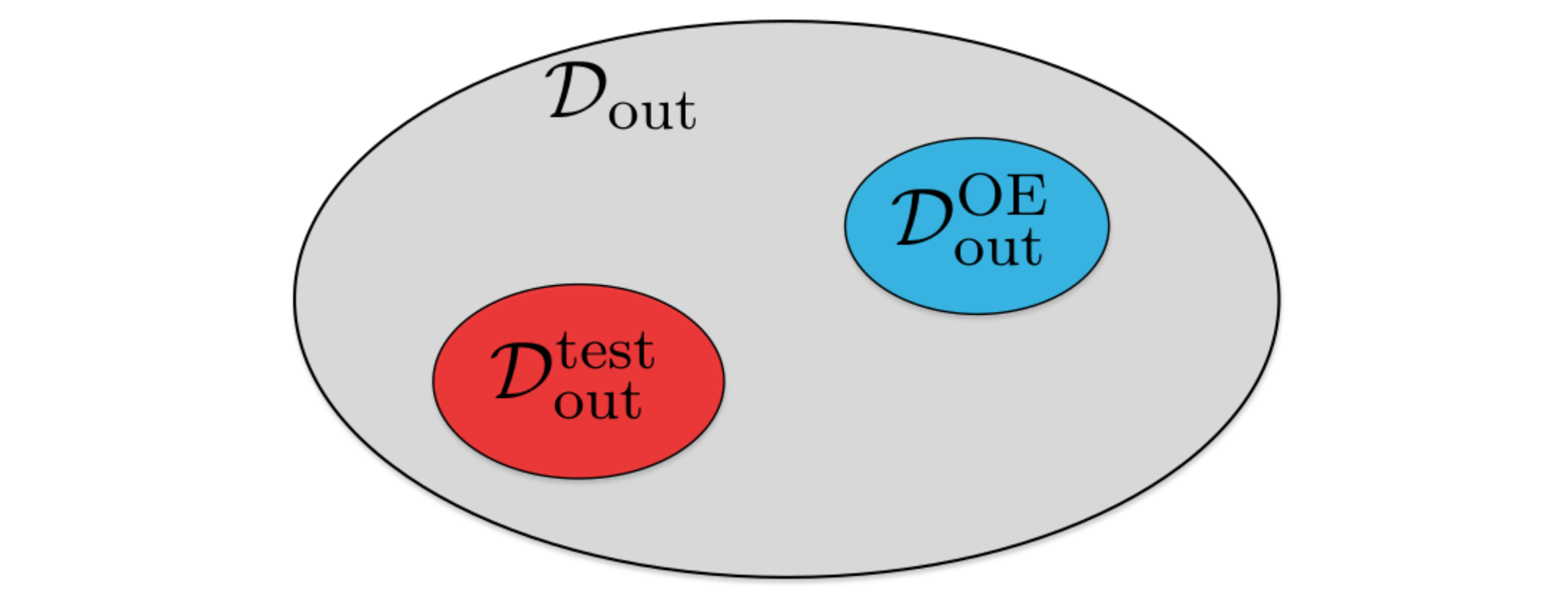}
         \caption[]{Illustrating the various OOD notations used throughout the text.}
         \label{fig:Dout}
    \end{center}
\end{figure}
Note that the complete set of out-of-distribution samples is represented by $\mathcal{D}_{\textrm{out}}$,
which is difficult or impossible to know \emph{a priori}.  When applying OE to training procedures, we are
intending to parametrize a subset of $\mathcal{D}_{\textrm{out}}$, labeled as 
$\mathcal{D}_{\textrm{out}}^{\textrm{OE}}$. During testing, another subset of OOD samples that are 
different from $\mathcal{D}_{\textrm{out}}^{\textrm{OE}}$ is introduced, and is defined to be 
$\mathcal{D}_{\textrm{out}}^{\textrm{test}}$. The nuances in notation are illustrated
by \figref*{fig:Dout}.

\subsubsection{Signal Model} \label{subsect:SignalModel}
Examples exhibiting weak interference still primarily represent valid classes. As such, the use of disjoint data sets
in \citeref*{hendrycks2019deep}, or the sample replacement strategy in \citeref*{ren2019likelihood}, are not
applicable here. We therefore need to define OE for OFDM signals experiencing co-channel interference. 

The continuous time OFDM signal $r(t)$ is described by
\begin{align}
    r(t) &= \mathcal{A} 
          \sum_{k=0}^{K-1} \sum_{-\infty}^{\infty} s_{k,l} e^{j 2 \pi k \delta f_k (t - lT - \epsilon T)} g(t-lT-\epsilon T) 
          \nonumber \\
          &+ n(t),
    \label{eqn:rt}
\end{align}
where $\mathcal{A}=a e^{j \theta} e^{j 2 \pi \delta f_c t}$, $K$ is the number of subcarriers, $\delta f_k$ is the subcarrier spacing, $T$ is the OFDM symbol period, 
and $s_{k,l}$ is the modulation symbol value for the $k_{th}$ subcarrier of the $l_{th}$ OFDM symbol. 
The function $g(t)$ is the pulse shape resulting from any transmit or receive filtering, 
and $n(t)$ represents zero-mean, complex additive Gaussian noise.
$a$, $\delta f_c$, and $\epsilon$ are the power factor, carrier frequency offset, and
time offset, respectively. In the presence of co-channel interference $I(t)$,
~\eqnref*{eqn:rt} may be rewritten as
\begin{align} 
    r_{\textrm{OE}}(t) &= \mathcal{A}
          \sum_{k=0}^{K-1} \sum_{-\infty}^{\infty} s_{k,l} e^{j 2 \pi k \delta f_k (t - lT - \epsilon T)} g(t-lT-\epsilon T) \nonumber \\
          &+ n(t) + I(t),
    \label{eqn:rt_rewrite}
\end{align}
where $\mathcal{A}=a e^{j \theta} e^{j 2 \pi \delta f_c t}$. The goal now is to select a model for $I(t)$, which will serve as $\mathcal{D}_{\textrm{out}}^{\textrm{OE}}$ for OE
purposes.

Recall our constraint of making no assumptions about the types of interference encountered; we desire our OE to 
generalize to a broad class of co-channel interference. For inspiration, we look to the OFDM 
jammer effectiveness models discussed in \citeref*{shim2009}. There, Luo \emph{et al}. compare the effect of multiple interference
models on OFDM receiver BER under varying channel conditions and SIR.
Their conclusion was that multi-tone interference (MTI) was more effective than either barrage noise interference (BNI)
or partial band interference (PBI) under additive white Gaussian noise (AWGN). Furthermore, MTI makes no attempt to
model a specific waveform or protocol, and provides enough configuration flexibility to model both co-channel and
adjacent channel interference. We therefore model I(t) as 
\begin{align}
    I(t) = a_i \sum_{j=1}^{J} e^{2 \pi f_j t + \phi_j},
\end{align}
where $a_i$ is the power factor, $f_j$ is the frequency of the $j_{th}$ tone, and $\phi_j$ is the phase of the $j_{th}$ tone 
drawn from a uniform distribution over $[0,2 \pi]$. 

The fraction of tones overlapping OFDM subcarriers is defined as $\rho = q/K$,
where $q$ is drawn from a uniform distribution over $[\frac{K}{8}, K]$; these were chosen to represent the continuum 
between partial and full overlap with each OFDM symbol in the spectrum. Note that we have no requirement for
$J$ to be a contiguous set of integers with respect to $K$ or otherwise.

\subsection{Maximum Softmax Probability \label{sect:MSP}}
As discussed above, several works derive an OOD detection capability from a primary task, such as classification. MSP, developed by Hendrycks and Gimpel, is an effective OOD baseline built on top of
classification models \citep{hendrycks2018baseline}. Specifically, this approach utilizes the predicted (\emph{i.e.}, maximum) softmax score to determine if a given example is ID or OOD. The authors
demonstrate that while deep neural networks are over-confident in their predictions, even those generated from OOD data, the scores from OOD data are generally lower than scores from ID data. This enables the detection of OOD data via a surprisingly simple, but effective method: just compare the softmax score of a given example to prediction statistics generated from a set of known ID data.

Unlike many of the classification-based OOD detection works, our primary goal is to determine if an example is ID or OOD, not to classify it. However, given that MSP is built on top of a classification
model, we must create an auxiliary classification task to enable its use. We take the approach of classifying the (4) OFDM physical layer subcarrier modulation types discussed in 
\sectref*{sect:ExpSetup}. However, given the strong similarity between ID and OOD signal traffic, we add outlier exposure to make this approach more sensitive to weak interferers.

\subsubsection{Outlier Exposure}
We leverage the OE approach for multiclass classification as described in \citeref*{hendrycks2019deep}, where $\mathcal{L}_{\textrm{OE}}$ 
is defined as the cross-entropy between the softmax scores of
the outlier data and a uniform distribution. Theoretically (and as we demonstrate below, empirically), this results in the model being 
less confident when classifying signals with weak interferers present.

\subsubsection{Implementation}
For experimentation purposes, we utilize a modified pre-activation ResNet-34~\citep{He2016IdentityMI,He2016} architecture, 
which we describe here. First, all 2D convolutions are replaced with 1D
convolutions to handle the shape of the signal data.  Second, similar to \citeref*{Zagoruyko2016WideRN},
dropout~\cite{Srivastava2014DropoutAS} is included in each residual block right before the second convolutional layer.
Lastly, all strided convolutions are removed and replaced with dilated convolutions~\cite{Yu2016}. We take inspiration from the approach
described in \citeref*{Oord2016WaveNetAG} and increase the dilation factor by a power of two with each residual block 
up to a factor of 128 before repeating. 

\subsection{Distance Metric Learning \label{sect:DML}}
DML attempts to learn a distance consistent with semantic similarity~\cite{movshovitzattias2017fuss}. In other words,
a network is trained to generate representations in an embedding space, where inputs drawn from the same class or
distribution are close to each other, but also separated from those drawn from a different class or distribution. 
The relevant task here is to learn an embedding representing normal network traffic, which is separable from that experiencing weak co-channel interference. \citeref*{Schroff_2015} introduced the triplet loss, which optimized the relative distance between
select triples consisting of an anchor and both positive (similar) and negative (dissimilar) examples.
\citeref*{movshovitzattias2017fuss} introduced the proxy-NCA loss, which improved upon triplet-loss performance and
convergence rates by utilizing proxies for entire subsets of training data.
We utilize the proxy-anchor loss~\cite{kim2020proxy}, which takes advantage of both the data-to-data relations
described by the triplet loss and the data-to-proxy relations described by the proxy-NCA loss. 

The proxy-anchor loss is defined as
\begin{align}
    \ell (X) &= \frac{1}{|P^+|} \sum_{p \,\in P^+} \log \left(1 + \sum_{x \,\in X_p^+} e^{-\alpha (s(x,p) - \delta)} \right) \nonumber \\
             &+ \frac{1}{|P|} \sum_{p \,\in P}  \log \left(1 + \sum_{x \,\in X_p^-} e^{\alpha(s(x,p)+\delta)}\right), 
             \label{eqn:proxyloss}
\end{align}
where $P$ is the set of all proxies, $P^+$ is the set of positive
proxies within a given batch of data, $X_p^+$ is the set of positive embedding vectors for proxy $p$, and $X_p^-$ is the 
disjoint set $X-X_p^+$. Using the cosine similarity $s$ between embeddings $x$ and $p$, $\ell$ decreases when $x \in X_p^+$ is close to $p \in P^+$, and increases otherwise. As a result, $p$ and its positive examples are pulled closer together, 
while $p$ and its negative examples are pushed further apart. The proxies are learned while
training~\cite{movshovitzattias2017fuss}, and $\alpha$ and $\delta$ are used for scaling and margin, respectively,

Using proxy-anchor loss, we can train a network to learn unique embeddings representing normal network traffic for each
of the valid subcarrier modulation types. This becomes a classifier-based training approach, where the (4) classes are defined in \tblref*{tab:Mod_SINR}. However, given the relatively weak interference levels we expect to identify for this
application, we utilize OE during training to make the trained network sensitive to weak interferers.  

\subsubsection{Outlier Exposure} 
For our application, we seek to learn unique embeddings for the (4) classes representing valid subcarrier 
modulation types and the (4) classes representing valid subcarrier modulation types experiencing weak interference. 
This results in (8) classes total, with a single classifier-based learning objective utilizing the proxy-anchor loss. 
Therefore,~\eqnref*{eqn:OE_loss} may be reformulated as
\begin{align}
    \mathbb{E}_{(x,y) \sim D_{\textrm{in}}, D_{\textrm{out}}^{\textrm{OE}}} \left[ \mathcal{L}(f(x), y) \right],
\end{align}
where $\mathcal{L}$ is defined by \eqnref*{eqn:proxyloss}. 

\subsubsection{OOD Score}
\begin{figure}[t!]
\begin{center}
\includegraphics[scale=0.45]{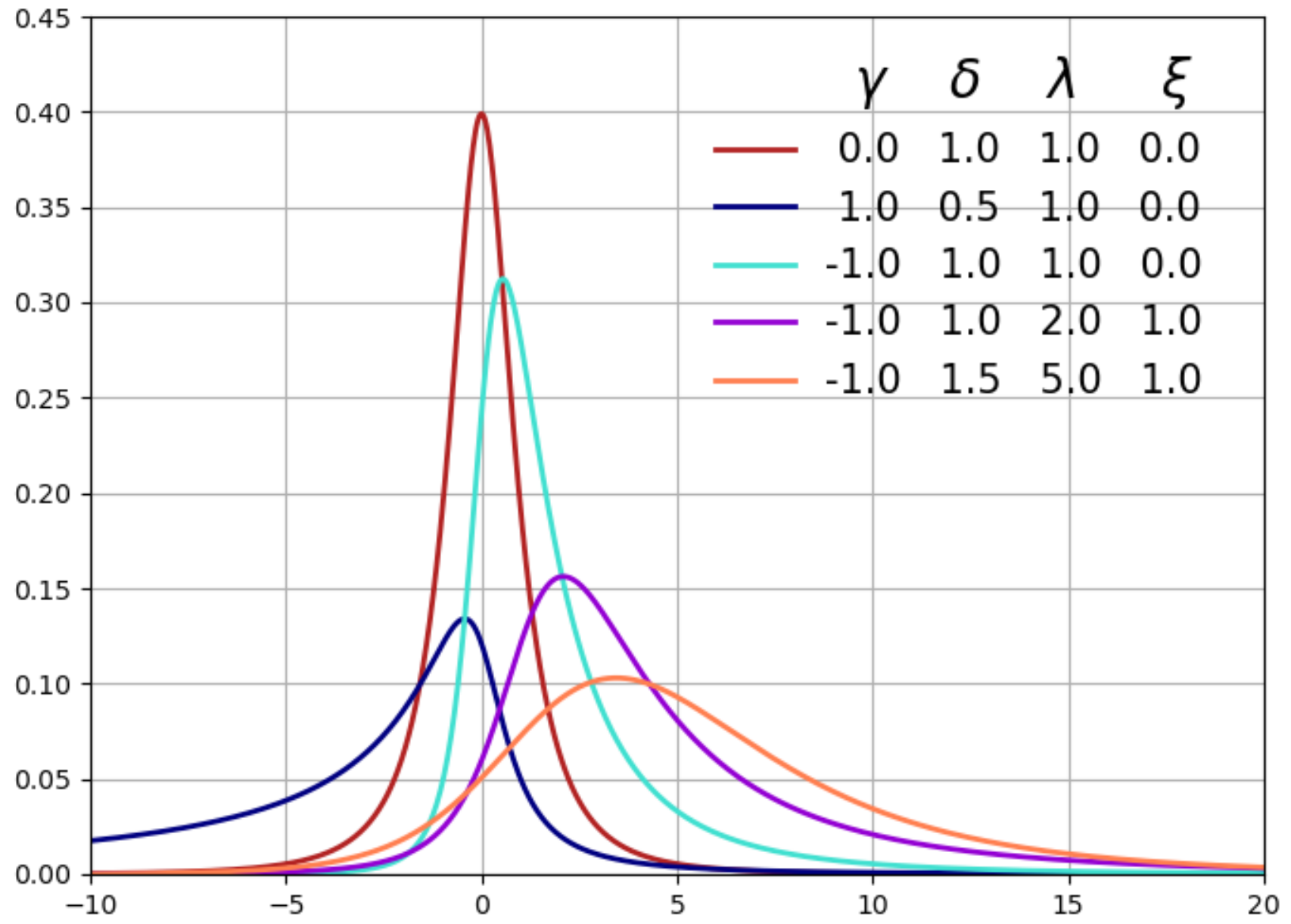}
 \caption[]{Various examples of the Johnson SU PDF.}
 \label{fig:JohnsonSU}
\end{center}
\end{figure}

An OOD metric is based on the distance between embeddings representing normal network traffic and those 
experiencing weak interference. Each embedding $\in \mathbb{R}^d$, where $d$ is the number of dimensions, \emph{e.g.}
64, 128, 256, etc. Many distance metrics however lose meaning as $d$ increases. For example, \citeref*{Aggarwal2001}
attempts to mitigate this limitation through the use of fractional distance metrics.
Cosine similarity does not have the same limitation, but substantial information may be lost when calculating the similarity 
between each embedding and its mean values only. We therefore look to model the probability density function (PDF) of each 
dimension in order to fully characterize normal behavior. Since each dimension has no incentive to strictly follow a normal 
distribution, each dimension is modeled as an unbounded Johnson distribution (Johnson SU) with the following PDF:
\begin{align}
    f(x) = \frac{\delta}{\lambda \sqrt{2 \pi} \sqrt{z^2 + 1}} e^{-\frac{1}{2} (\gamma+\delta \sinh^{-1} z)^2},
\end{align}
where $z$ is parametrized as $\frac{x - \xi}{\lambda}$, $\xi$ is a location parameter, $\lambda$ is a scaling parameter,
and $\gamma$ and $\delta$ are shape parameters~\cite{Cayton2015}. The Johnson SU
PDF is capable of modeling various asymmetric distributions; see \figref*{fig:JohnsonSU},
where each curve represents the indicated $(\gamma, \delta, \lambda, \xi)$ tuple.

\subsubsection{Implementation}
For experimentation purposes, we utilize a ResNet-34~\cite{He2016} architecture, but with the convolution operations 
replaced by the dilated convolution operation~\cite{Yu2016}. These expand the receptive field without 
loss of resolution. Since we are operating on complex data, the I and Q components of each sample are separated and placed in separate input channels, similar to the separate R, G, and B
channels utilized for image processing. The ResNet output is pooled and flattened, and a dropout of 0.25 applied. 
The result is then run through a dense layer with an output width equal to our desired embedding size of 128. Training is performed using AdamW~\cite{Loshchilov2019} optimization and a cosine annealing~\cite{Loshchilov2017} period of 10.

\subsection{Variational Autoencoder \label{sect:VAE}}
A variational autoencoder (VAE)~\cite{kingma_2013} is a popular unsupervised learning technique, and is  
capable of generating or reconstructing examples similar to the input. The model is trained on  
signals that are considered normal or expected. In inference mode, when the model is given a 
signal experiencing a weak interferer, the reconstruction should be relatively poor, allowing 
for a method of detection. The loss metric for a traditional autoencoder is often 
the mean squared error between the input $x$ and the reconstructed output $\hat{x}$. 
The variational approach, however, encodes the inputs as probability
distributions, which regularizes the latent space by constraining each dimension to be approximately
Gaussian. The decoder then samples the latent space to generate new noisy examples; sampling is used 
to overcome the overfitting nature of conventional autoencoders. The mathematical details
of VAEs are summarized below; see Refs.~\cite{kingma_2013, rezende_2014, doersch2016tutorial, Blei_2017,cheng2021variational} for more
information.

A VAE is a continuous latent variable model. An unobserved latent 
representation $z$ is used to generate an observation $x$. The probability distribution 
$p_{\theta}(x|z)$, parameterized by $\theta$, describes the process of generating
$x$ from a lower-dimensional representation $z$ described by prior $p(z)$. Assuming 
the latent probability distribution $p(z)$ is known, the goal is to learn
the marginal likelihood of the data:
\begin{align}
     \max_{\phi, \theta} \mathbb{E}_{q_{\phi}(z|x)} \left[ \log p_{\theta}(x|z)\right],
\end{align}
which may be rewritten as
\begin{align}
    \log p_\theta(x|z) = D_{\textrm{KL}} (q(z|x) \,\, || \,\, p(z))  + \mathcal{L}(\theta, \phi; x, z),
\end{align}
where $\phi, \theta$ parameterize the the encoder and decoder distributions respectively.
The $D_{\textrm{KL}}(||)$ term is the non-negative Kullback-Leibler (KL) divergence between the true and approximate 
posterior. To make the optimization tractable, $ \mathcal{L}(\theta, \phi; x, z)$ is maximized:
\begin{align}
     \log p_\theta(x|z) \geq \mathcal{L}(\theta, \phi; x, z) =
     \mathbb{E}_{q_{\phi}(z|x)}\left[ \log p_{\theta}(x|z)\right] \nonumber \\
     - D_{\textrm{KL}} \left(q(z|x) || p(z)\right)
\end{align}
where the prior $p(z)$ and posterior $q_\phi(z|x)$ are approximated as Gaussian.
Rewritten more simply, the VAE loss is
\begin{align}
     \mathcal{L}_{\textrm{VAE}} = \mathcal{L}_{\textrm{MSE}} + \beta \mathcal{L}_{\textrm{KL}},
     \label{eqn:LVAE}
\end{align}
where $\mathcal{L}_{\textrm{MSE}} \equiv \mathbb{E}_{q_{\phi}(z|x)}\left[ \log p_{\theta}(x|z)\right]$ is
commonly taken as the mean squared error between the input $x$ and the output $\hat{x}$, and $\mathcal{L}_{\textrm{KL}}$
has an analytical form under Gaussian assumptions. The empirical factor $\beta$ 
is often introduced to scale $\mathcal{L}_{\textrm{KL}}$, referred to as a $\beta$-VAE. As $\beta \rightarrow 0$,
the VAE reduces to the traditional deterministic autoencoder. Increasing the value of $\beta$ has been motivated
to further constrain the capacity of the encoder in order to allow a more factorized latent
representation~\cite{higgins_2016,burgess2018understanding}. 

\subsubsection{Outlier Exposure}
The goal is to minimize $\mathcal{L}_{\textrm{VAE}}$ using
ID data while simultaneously maximizing a loss associated with outlier exposure, defined to be
$\mathcal{L}_{\textrm{OE}}$, which utilizes both ID and OOD data. OE is incorporated
into~\eqnref*{eqn:LVAE} as a penalty term:
\begin{align}
    \mathbb{E}_{x \sim \mathcal{D}_{\textrm{in}}} \mathcal{L}_{\textrm{VAE}}(x) +
    \lambda \mathbb{E}_{x \sim \mathcal{D}_{\textrm{in}}, \, x' \sim \mathcal{D}_{\textrm{out}}^{\textrm{OE}}}
    \mathcal{L}_{\textrm{OE}}(x, x'), 
\end{align}
where $\lambda$ is a scaling factor, $\mathcal{D}_{\textrm{in}}$ represents the ID dataset, 
and $\mathcal{D}_{\textrm{out}}^{\textrm{OE}}$ represents the OOD dataset used during training. 
The loss function with the OE modification is
\begin{align}
    \mathcal{L} = \mathcal{L}_{\textrm{VAE}} - \lambda \mathcal{L}_{\textrm{OE}},
    \label{eqn:VAE_Loss}
\end{align}
where $\mathcal{L}_{\textrm{OE}}$ may be calculated using the approaches described
in \citeref*{cheng2021variational}. We choose to calculate $\mathcal{L}_{\textrm{OE}}$
using the input $x$ and the corresponding reconstructed output $\hat{x}$:
\begin{align}
            \mathcal{L}_{\textrm{OE}} = \sigma \left( \textrm{MSE}_{\textrm{OOD}}(x', \hat{x}') - 
                                        \textrm{MSE}_{\textrm{ID}}(x, \hat{x})\right),
            \label{eqn:OE_MSE}
        \end{align}
where $\sigma$ is the typical sigmoid function, and $x$ and $x'$ live in $\mathcal{D}_{\textrm{in}}$ and 
$\mathcal{D}_{\textrm{out}}^{\textrm{OE}}$, respectively.
\eqnref*{eqn:VAE_Loss} wants to reconstruct ID data well, while poorly reconstructing OOD data.

\subsubsection{Implementation}
The encoder is used to build two latent representations, one each for I and Q separately. Then, two identical
decoding networks are used to regenerate both I and Q, which are then concatenated to reconstruct the complex input. 
The encoder block is constructed out of strided convolutional layers followed by batch normalization and
a ReLU activation function. It is relatively shallow, built from only two encoding blocks, which effectively
reduces the IQ input length by a factor of four while increasing the channel dimension by a factor of four.
A flattening operation is then performed to create the latent linear layer, which funnels the encoding features
from 512 to an empirically determined size of 128. The decoder is symmetric to the encoder, but uses strided transpose convolutional layers.

\subsection{Autoregressive Log Likelihood \label{sect:AR}}
Deep autoregressive models are a form of generative model that learn a distribution over a given input space. Unlike many other generative models,
\emph{e.g.}, GANs and VAEs, autoregressive models produce a tractable likelihood for a given input. Specifically, the joint probability of each 
input $\boldsymbol{x} = \{x_1,...,x_T\}$ is factorized into a product of conditional probabilities:
\begin{align}
    p(\boldsymbol{x}) = \prod_{t=1}^{T} p(x_t|x_1,...,x_{t-1}).
\end{align}
Stated another way, each input element $x_t$ is conditioned on all data from previous time steps. Traditionally, these conditional probability 
distributions are modeled using either recurrent neural networks (\emph{e.g.}, PixelRNN~\cite{Oord2016PixelRN}) or convolutional neural networks
(\emph{e.g.}, PixelCNN, WaveNet~\cite{Oord2016PixelRN,Oord2016WaveNetAG}), the latter trading an unbounded receptive field for reduced computation
cost at training due to parallelization. 

Using generative models for anomaly detection is a well established approach~\cite{Bishop1994NoveltyDA}, and deep autoregressive models are 
especially appealing for anomaly detection since they can tractably compute the likelihood of complex inputs. Similar to the MSP method, this
theoretically\footnote{Recently, Nalisnick \emph{et al.} demonstrated that deep generative models often assign higher likelihoods to OOD data
compared to ID data~\citep{Nalisnick2019DoDG}; however, we find our model does not suffer from this problem. We hypothesize this may due to the
similarity of our ID and OOD data.} facilitates the detection of OOD data by just comparing the computed likelihood of a given example to 
likelihood statistics generated from a set of known ID data. As in our other methods, we make use of outlier exposure to make this approach 
more sensitive to weak interferers.

\subsubsection{Outlier Exposure}
We leverage the OE approach for density estimation as described in \citeref*{hendrycks2019deep}, where $\mathcal{L}_{\textrm{OE}}$ is defined
as the margin loss over the log-likelihood difference between ID and OOD data:
\begin{align}
    \mathcal{L}_\textrm{OE} = \max(0, m + \textrm{NLL}(f_{\theta}(x)) - \textrm{NLL}(f_{\theta}(x'))),
\end{align}
where $f_{\theta}$ is a neural network with parameters $\theta$, NLL is the negative log-likelihood, $x$ and $x'$ are 
ID and OOD examples respectively. $m$ is a margin hyperparameter.

\subsubsection{Implementation}
The autoregressive model is based on temporal convolutional networks (TCNs), a family of architectures for sequence modeling tasks~\cite{Bai2018AnEE}.
TCNs are inspired by the WaveNet architecture~\cite{Oord2016WaveNetAG} while being less complex to implement. Like WaveNet, they make use of 
appropriate zero padding and causal convolutions to ensure the model is autoregressive in nature. In this work, we eschew the original architecture
developed in \citeref*{Bai2018AnEE} and instead use building blocks common to our other methods to enable a fairer comparison. Specifically, 
the 1D ResNet-34 architecture developed in the MSP and DML methods is used as the TCN backbone, modifying all convolutions to be 
causal. Note that this backbone is not enough by itself, as we must model the conditional probabilities $p(x_t|x_1,...,x_{t-1})$. To accomplish this,
the backbone is connected to a mixture density network (MDN)~\cite{Bishop1994MixtureDN}, which models the conditional distributions as a mixture of 
Gaussians. Note that this is in contrast to the PixelRNN, PixelCNN, and WaveNet architectures, which model the conditional probabilities using a softmax
distribution. Through testing, we found that the MDN improved detection performance while providing faster training and inference speed.

\section{Experiments \label{sect:Experiments}}
\newcommand{\douttest}{\mathcal{D}_{\textrm{out}}^{\textrm{test}}}
\newcommand{\doutoe}{\mathcal{D}_{\textrm{out}}^{\textrm{OE}}}
\newcommand{\dout}{\mathcal{D}_{\textrm{out}}}
\newcommand{\din}{\mathcal{D}_{\textrm{in}}}

\subsection{Signal Synthesis}

The samples generated for each experiment follow the scenario described in \sectref*{sect:ExpSetup} and the signal processing chain described in \sectref*{sect:DL_Approaches}.  The processing chain, including the OOD detection sub-system, is described in \figref*{fig:OOD}. ID samples are generated by following the 802.11 OFDM specification, and represent the nominal case when an aggressor is not currently interfering with the victim STA. At the point in the receive chain where OOD detection takes place, we assume coarse synchronization and equalization have already been achieved; therefore, synchronization error is not considered in this study.

OOD samples are generated in a similar manner as ID samples, except that a weak interfering signal is also added. We test with two different sources of co-channel interference; another 802.11 OFDM signal and an 802.11 DSSS signal. The OFDM interferer is identical in structure to the victim STA OFDM signal described by \eqnref*{eqn:rt}. The DSSS interferer is a DBPSK DSSS signal carrying a randomly generated payload, and is described as 
\begin{align}
    r_{\text{DSSS}} (t)=ae^{j\theta} e^{j2\pi\delta f_c t} s_p x_m g(t-mT),
    \label{eqn:DSSS}
\end{align}
where $s_p$ represents a pseudo-random spreading sequence of length $P$; the duration of each value is $T/P$ seconds and can take values of $\pm{1}$~sec. 
The term $x_m$ represents each DBPSK symbol of duration of $T$ seconds, which also takes on values of $\pm{1}$~sec. 
$g(t)$ represents the impulse response of any receive or transmit filtering, 
$\theta$ is a constant phase offset in the range $[0,2\pi)$, $\delta f_c$ is the carrier frequency offset, and $a$ is a power factor.

As discussed in \sectref*{sect:ExpSetup}, the victim STA subcarrier modulation schemes are selected from \tblref*{tab:Mod_SINR}, which also defines the required SINR. An appropriate SIR is then selected and applied, and the remaining signal power is consumed by SNR. Additionally, the victim and aggressor STAs are not co-located, meaning that each signal experiences different channel effects. Assuming channel equalization in the receiver, corrections made to the desired OFDM signal result in further degradation to the interfering signal. An indoor WLAN TGn channel is used to model degradation to the aggressor STA signal at the receiver.

\subsection{Dataset Generation}
Two of the four datasets described in \sectref*{subsect:OE} are used to evaluate the performance of our approaches to weak interferer detection: $\din$ and $\douttest$. 
$\din$ is used as an ID reference, and the corresponding output considered the nominal ID representation. $\douttest$ contains examples with interferers
at varying SIR values. To evaluate the ability of each model to generalize across different types of interferers, each $\douttest$ contains
either an 802.11 OFDM interferer or an 802.11 DSSS interferer. To better understand how channel effects to each interferer affects model 
performance, four datasets are used, and include examples with and without channel effects; see \tblref*{tab:test_datasets} for a summary.

\begin{table}[t!]
\caption[]{OOD test dataset notations used during experimentation.}
\centering
    \begin{tabular}{c c c} \toprule
      \textbf{Dataset} & \textbf{Interferer} & \textbf{Apply Channel Effects to Interferer} \\ \midrule
        $(\douttest)_1$ &	DSSS	& No  \\
        $(\douttest)_2$ &	DSSS	& Yes \\
        $(\douttest)_3$ &	OFDM	& No  \\
        $(\douttest)_4$ &	OFDM    & Yes \\
        \bottomrule
        \hline
    \end{tabular}
\label{tab:test_datasets}
\end{table}

In addition to varying interferer types, detector performance is evaluated with and without OE. Networks trained with OE require additional datasets,
labeled $\doutoe$, which represent cases where parameterized co-channel interference is present;  \sectref*{subsect:SignalModel} provides a detailed
description of the MTI-based interferers used during training. Note that the goal is for $\doutoe$ to generalize to $(\douttest)_n$, even though the datasets 
have distinct signal models sharing minimal physical structure.

The datasets are formatted to have the following dimensions: 
\begin{gather*}
    [\textrm{\footnotesize{Dataset dimensions}}] 
    \times 
    [\textrm{\footnotesize{Input sample dimensions}}] = \nonumber \\
    [\textrm{\footnotesize{n\_mod}}, \textrm{\footnotesize{n\_sir\_bins}}, \textrm{\footnotesize{n\_batches}},\textrm{\footnotesize{batch\_size}}] 
    \times 
    [\textrm{ \footnotesize{block\_size}}, \textrm{\footnotesize{n\_channels}}], \nonumber
\end{gather*}
where the n\_mod dimension supports detector characterization for specific MCS values, allowing us to focus on model performance across various SIR values. To provide detailed detector performance across SIR values, SIR is binned into (14) equi-spaced linear values, annotated by n\_sir\_bins.  The variables 
n\_batches and batch\_size represent the number of batches per dataset and number of items per batch, respectively. Lastly, block\_size represents the number of complex samples comprising each example, and n\_channels are the real and imaginary components of the complex data.

\begin{table}[b!]
\caption[]{Definitions of dataset dimensions.}
\centering
    \begin{tabular}{l c l} \toprule
      \textbf{Name} & \textbf{Value} & \textbf{Description}               \\ \midrule
        n\_mod	     &  4	&	number of mod. schemes used by victim STA \\
        n\_sir\_bins & 	14	&	number of SIR bins for plotting           \\
        n\_batches	 &	16	&	number of batches per SIR bin             \\
        batch\_size	 &	64	&	number of items in a batch                \\
        \hline \\
        block\_size	 &	960	&	number of complex signal samples considered \\
        n\_channels	 &	2	&	number of channels (real and imaginary)     \\
        \bottomrule
        \hline
    \end{tabular}
\label{tab:dataset_dimensions}
\end{table}

\subsection{Inference}
To generate a comprehensive set of performance curves, inference results are obtained for all combinations of models and datasets $(\douttest)_n$. 
Note that all models take input samples of size $[960,2]$ and produce a single inference output score, which represents a metric that is used to flag
OOD samples. The eight models used for inference are the four architectures described in this paper, trained both with and without OE. 
The trained models are listed in \tblref*{tab:model_details}. 

\begin{table}[t!]
\caption[]{The models that have been evaluated.}
\centering
    \begin{tabular}{l c c r} \toprule
        \textbf{Model} & \textbf{Section} & \textbf{OE Training} & \textbf{\# Parameters} \\ \midrule
        AR          & \ref{sect:AR}  & No   & 7556k \\ 
        AR, OE      & \ref{sect:AR}  & Yes  & 7556k \\ 
        DML         & \ref{sect:DML} & No   & 7250k \\ 
        DML, OE     & \ref{sect:DML} & Yes  & 7250k \\ 
        MSP         & \ref{sect:MSP} & No   & 7219k \\ 
        MSP, OE     & \ref{sect:MSP} & Yes  & 7219k \\ 
        VAE         & \ref{sect:VAE} & No   & 278k  \\ 
        VAE, OE     & \ref{sect:VAE} & Yes  & 278k  \\ 
        \bottomrule
        \hline
    \end{tabular}
\label{tab:model_details}
\end{table}

\begin{figure}[h!]
\begin{center}
    \includegraphics[scale=0.4]{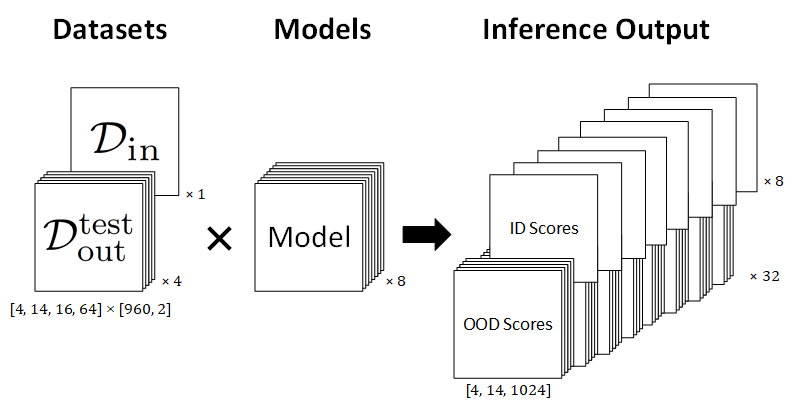}
\caption[]{Combining datasets and models from Tables~\ref{tab:test_datasets} and \ref{tab:model_details} for inference.}
\label{fig:inference_methodology}
\end{center}
\end{figure}

For each of the (8) trained models, the $\din$ and each of the four $\douttest$ datasets is provided as input to
produce a total of (5) sets of inference output scores per model: (1) set of ID scores and (4) sets of OOD scores.
Since there are a total of $4 \times 14 \times 1024$ input samples per dataset, there are also a total of 
$4 \times 14 \times 1024$ inference output scores per dataset. \figref*{fig:inference_methodology} shows the 
expansion of inference output data for all combinations of models and datasets.

\subsection{Evaluation}
We define an experiment as the evaluation of performance metrics given a choice of model from \tblref*{tab:model_details} 
and $\douttest$ from \tblref*{tab:test_datasets}. For each experiment, a set of both ID and OOD inference 
scores are generated, which are then combined to construct a receiver operating characteristic (ROC) curve. This is an appealing 
performance metric given that a threshold does not need to be specified, but can be chosen later for a specific operational use-case. 

A ROC curve is defined as the relationship between the true positive rate (TPR) and false positive rate (FPR) over 
a varying threshold. TPR and FPR are calculated as
\begin{align}
    \textrm{TPR} = \frac{\textrm{TP}}{\textrm{TP} + \textrm{FP}}, \hspace{0.75cm} \textrm{FPR} = \frac{\textrm{FP}}{\textrm{FN} + \textrm{TN}},
    \label{eqn:TPR}
\end{align}
where TP, TN, FP, FN are the number of true positives, true negatives, 
false positives, and false negatives, respectively. 
We define OOD and ID samples to be the positive and negative classes respectively. 
Since we have discretized the SIR range into 14 bins, there are 14 ROC curves per modulation experiment, where each curve
is constructed from 1024 ID and 1024 OOD samples.

\begin{figure}[b!]
\begin{center}
    \includegraphics[scale=0.55]{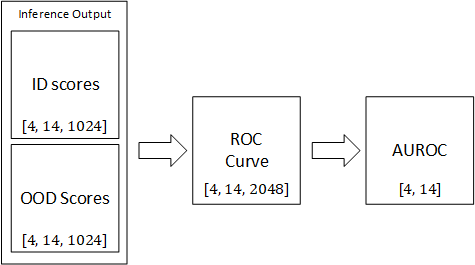}
\caption[]{Combining inference scores in order to evaluate an experiment.}
\label{fig:evaluation_methodology}
\end{center}
\end{figure}

The area under the ROC curve (AUROC) metric is used in order to reduce each curve to a scalar  
performance indicator for ease of comparison across experiments and SIR bins. Since there are $4 \times 14$ ROC curves generated
for each experiment, one for each subcarrier modulation type and SIR bin, there are also $4 \times 14$ AUROC scores for
each experiment. This is further illustrated in \figref*{fig:evaluation_methodology}.

\section{Results \label{sect:Results}}
\begin{figure*}[t!]
    \begin{center}
        \includegraphics[scale=0.4]{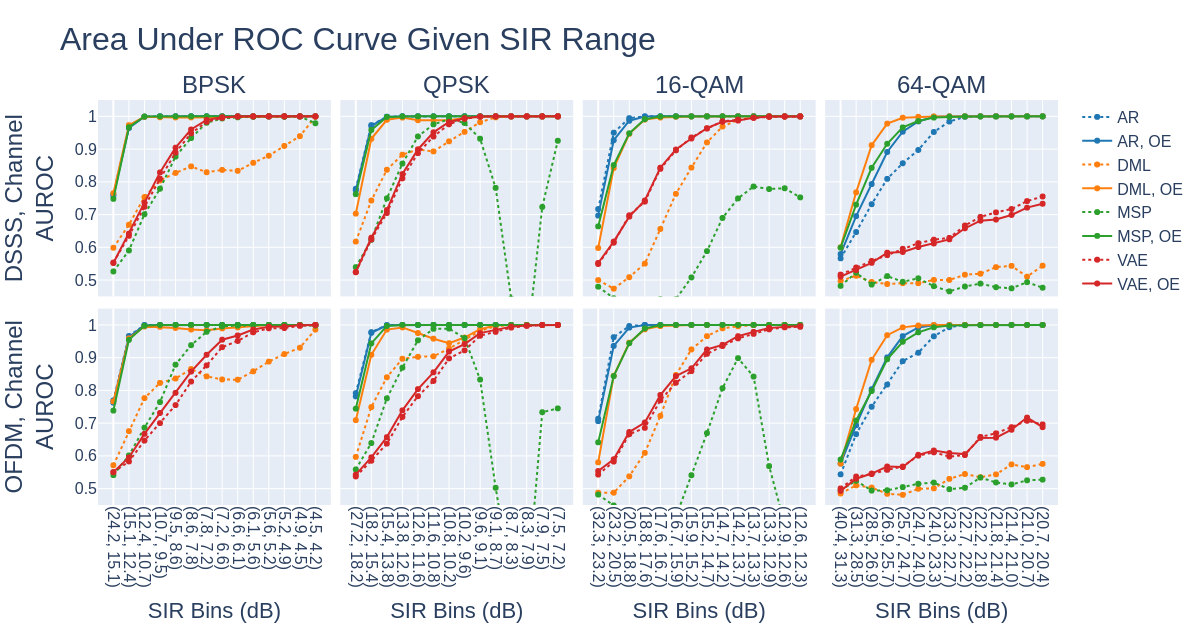}
        \caption[]{AUROC scores over SIR bins (dB) for all models and datasets. SIR bins are displayed in descending order to show increase in aggressor STA signal strength relative to victim STA signal strength.}
        \label{fig:auroc_datasets}
    \end{center}
\end{figure*}


\figref*{fig:auroc_datasets} summarizes the detection performance of each of the DL approaches discussed in \sectref*{sect:DL_Approaches}. Each graph plots AUROC scores vs. SIR bin values for one of the subcarrier modulation types listed in \tblref*{tab:Mod_SINR}. The datasets used to calculate AUROC scores were $(\mathcal{D}_{\textrm{out}}^{\textrm{test}})_2$ and
$(\mathcal{D}_{\textrm{out}}^{\textrm{test}})_4$. Note that each point on each plot represents the area under a single ROC curve generated from 2048 inference scores. Experiments on models trained with OE are plotted as solid lines, and those without OE plotted as dotted lines. 

Overall, the AR, DML, and MSP models all perform well with OE applied. For all subcarrier modulation types, each exhibit a monotonically increasing shape as SIR decreases. The DML model does deviate slightly for some QPSK SIR values, but exhibits the best overall performance for 64-QAM. This may be an indication that more than 2048 inference scores are required for each SIR value. The AR model exhibits the least difference in performance between its base model and that with OE applied. The VAE model clearly needs greater SIR levels to reach similar AUROC values, especially for higher constellation orders. It also appeared to benefit little from OE. However, given its smaller size as indicated in \tblref*{tab:model_details}, the VAE model may be an appealing approach for resource constrained platforms.   

Clearly the DML and MSP models perform much better in this application with OE applied, with the MSP model exhibiting the greatest benefit. Without it, \figref*{fig:auroc_datasets} shows two prominent peaks in MSP AUROC scores at various SIR values for both QPSK and 16-QAM subcarrier modulation types. We believe this to be caused by the network's failure at generalizing across perturbations caused by co-channel interference and overconfidence in classifications. Furthermore, this behavior seems to occur at lower SIR values, where perturbations would be at their greatest.

\section{Conclusion \label{sect:Conclusion}}
In this work we explored multiple DL approaches to persistent weak interference detection in WiFi networks. We also demonstrated a novel training approach whereby each model is taught the notion of outliers using a generalized signal interference model. The result was the ability to detect weak co-channel and adjacent channel interference at SIR levels at least 20~dB below  specified minimum sensitivity levels. As discussed in \citeref*{Gummadi-et-al-2007}, this capability has the potential to significantly improve both network throughput and latency. We also suspect this approach can be fielded at a fraction of the computational cost of current techniques, including SIC \cite{Souvik2010}, which is highly susceptible to estimation errors~\cite{Selim2019}~\cite{Huang2001}. We suggest exact computational costs of both approaches be the subject of future work in this area.


\appendix
\section*{Appendix A: Training Parameters \label{sect:AppendixA}}
All models are trained using the AdamW optimizer with batches of size 64. Two types of learning rate schedulers 
are used: cosine and cosine annealing with warm restarts. \tblref*{tab:common_training_params} shows the list of 
common training parameters, and \tblref*{tab:training_params} shows training parameters that varied over model 
architecture. Note that OE lambda and OE margin only apply for models trained with outlier exposure.

\begin{table}[t!]
\caption[]{Common training parameters.}
\centering
    \begin{tabular}{l c} \toprule
        Optimizer           & AdamW \\
        Weight Decay        & 0 \\
        LR                  & 0.001 \\ 
        Batch Size          & 64  \\ 
        \bottomrule
        \hline
    \end{tabular}
\label{tab:common_training_params}
\end{table}

\begin{table}[t!]
\caption[]{Model specific training parameters.}
\centering
    \begin{tabular}{l c c c c} \toprule
        \textbf{Parameter} & \textbf{AR} & \textbf{DML} & \textbf{MSP} & \textbf{VAE}\\ \midrule
        LR Scheduler        & Cosine & CosAnWR & Cosine & CosAnWR \\ 
        OE Lambda           & 1.0    & N/A     & 1.0    & 500.0\\ 
        OE Margin           & 1.0    & N/A     & N/A    & 2.0 \\ 
        Epochs              & 50     & 100     & 50     & 100  \\ 
        Batches per Epoch   & 1024   & 512     & 1024   & 65 \\ 
        \bottomrule
        \hline
    \end{tabular}
\label{tab:training_params}
\end{table}

\section*{Appendix B: Intermediate Results \label{sect:AppendixB}}
\begin{figure*}[p!]
\begin{center}
    \includegraphics[scale=0.6]{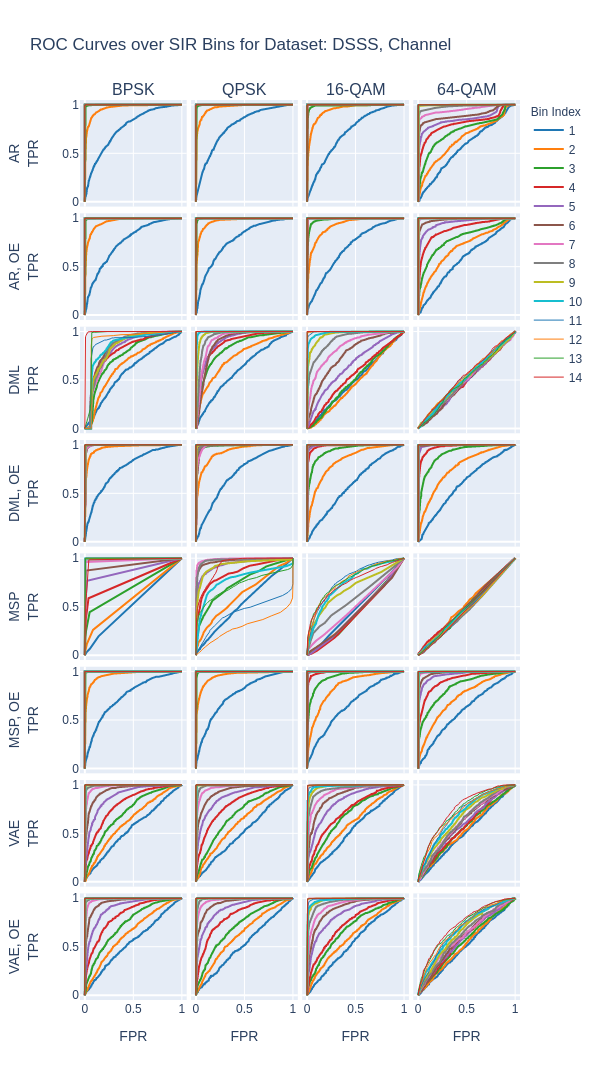}
    \captionsetup{justification=centering}
    \caption[]{All ROC curves for the DSSS weak interferer with channel effects.}
    \label{fig:sir_roc}
\end{center}
\end{figure*}

\figref*{fig:sir_roc} displays the complete set of ROC curves used to calculate the AUROC scores in the 
first column of \figref*{fig:auroc_datasets}, which corresponds to dataset 
$(\mathcal{D}_{\textrm{out}}^{\textrm{test}})_1$. The color of each ROC curve describes the SIR bin index. 
\tblref*{tab:bin_indices} provides the SIR range for each ROC curve given a bin index and subcarrier modulation type.


\begin{table}[ht!]
\caption[]{SIR bin ranges (dB) for \figref*{fig:sir_roc}.}
\centering
    \begin{tabular}{r c c c c} \toprule
        \textbf{Bin Index} & \textbf{BPSK} & \textbf{QPSK} & \textbf{16-QAM} & \textbf{64-QAM} \\
        & (max, min) & (max, min) & (max, min) & (max, min) \\ \midrule
        1  & (24.2, 15.1) & (27.2, 18.2) & (32.3, 23.2) & (40.4, 31.3) \\ 
        2  & (15.1, 12.4) & (18.2, 15.4) & (23.2, 20.5) & (31.3, 28.5) \\ 
        3  & (12.4, 10.7) & (15.4, 13.8) & (20.5, 18.8) & (28.5, 26.9) \\ 
        4  & (10.7, 9.5)  & (13.8, 12.6) & (18.8, 17.6) & (26.9, 25.7) \\ 
        5  & (9.5, 8.6)   & (12.6, 11.6) & (17.6, 16.7) & (25.7, 24.7) \\ 
        6  & (8.6, 7.8)   & (11.6, 10.8) & (16.7, 15.9) & (24.7, 24.0) \\ 
        7  & (7.8, 7.2)   & (10.8, 10.2) & (15.9, 15.2) & (24.0, 23.3) \\ 
        8  & (7.2, 6.6)   & (10.2, 9.6)  & (15.2, 14.7) & (23.3, 22.7) \\ 
        9  & (6.6, 6.1)   & (9.6, 9.1)   & (14.7, 14.2) & (22.7, 22.2) \\ 
        10 & (6.1, 5.6)   & (9.1, 8.7)   & (14.2, 13.7) & (22.2, 21.8) \\ 
        11 & (5.6, 5.2)   & (8.7, 8.3)   & (13.7, 13.3) & (21.8, 21.4) \\ 
        12 & (5.2, 4.9)   & (8.3, 7.9)   & (13.3, 12.9) & (21.4, 21.0) \\ 
        13 & (4.9, 4.5)   & (7.9, 7.5)   & (12.9, 12.6) & (21.0, 20.7) \\ 
        14 & (4.5, 4.2)   & (7.5, 7.2)   & (12.6, 12.3) & (20.7, 20.4) \\
        \bottomrule
        \hline
    \end{tabular}
\label{tab:bin_indices}
\end{table}

\section*{Acknowledgment \label{sect:Acknowledgment}}
We would like to thank Matthew Kinsey, Ryan Sawyer, Kate Tallaksen, and Richard Shaner for their valued contributions to the various research topics and modeling efforts comprising this work. 

\clearpage

\clearpage

\bibliographystyle{unsrt}
\bibliography{references}

\end{document}